\documentclass[twocolumn,showpacs,preprintnumbers,amsmath,amssymb,nofootinbib]{revtex4-2}
\usepackage{comment} 
\usepackage{cancel}
\usepackage{graphicx}
\usepackage{dcolumn}
\usepackage{bm}
\usepackage[compact]{titlesec}

\makeatletter
\DeclareRobustCommand{\cev}[1]{%
  {\mathpalette\do@cev{#1}}%
}
\newcommand{\do@cev}[2]{%
  \vbox{\offinterlineskip
    \sbox\z@{$\m@th#1 x$}%
    \ialign{##\cr
      \hidewidth\reflectbox{$\m@th#1\vec{}\mkern4mu$}\hidewidth\cr
      \noalign{\kern-\ht\z@}
      $\m@th#1#2$\cr
    }%
  }%
}
\makeatother
\usepackage{xcolor}

\usepackage{hyperref}
\renewcommand{\arraystretch}{2.3}       
\newcommand{\half}{{{\textstyle\frac{1}{2}}}}
\newcommand{\quarter}{{{\textstyle\frac{1}{4}}}}
\newcommand{\be}{\begin{equation}}
\newcommand{\ee}{\end{equation} }
\newcommand{\beqa}{\begin{eqnarray} }
\newcommand{\eeqa}{\end{eqnarray} }
\newcommand{\ba}{\begin{array}}
\newcommand{\ea}{\end{array}}
\newcommand{\bpm}{\begin{pmatrix}}
\newcommand{\epm}{\end{pmatrix}}

\newcommand{\dis}{\displaystyle}

\newcommand{\rmd}{{\rm d}}
\newcommand{\rd}{{\rmd}}

\newcommand{\ODD}{\mathbf{O}(D,D)}

\newcommand\To{T_{\scriptscriptstyle{{(0)}}}}

\newcommand\Tr{{\scalebox{0.9}{${\mathrm{Tr}}$}}}

\newcommand\cG{{\cal G}}

\newcommand\cL{{\cal L}}

\newcommand\cT{{\cal T}}

\def\na{\nabla}

\newcommand\deltaH{{\delta_{{{H\!{-{\rm{flux}}}}}}}}
\newcommand\wK{{w_{{{K}}}}}
\newcommand\wT{{w_{{{T}}}}}
\newcommand\kB{{k_{{\rm{B}}}}}
\newcommand\LambdaQCD{{\Lambda_{{\rm{QCD}}}}}

\newcommand{\there}{{{\scriptscriptstyle{\rm{there}}}}}

\newcommand\rr{{r}}

\newcommand{\trd}{{\bigtriangledown}}

\newcommand{\rstar}{{{r_{\star}}}}
\newcommand{\betappn}{\beta_{{{{\rm{PPN}}}}}}
\newcommand{\gammappn}{\gamma_{{{\rm{PPN}}}}}
\newcommand{\GN}{G_{{{{{\rm{N}}}}}}}

\newcommand{\oBD}{\scalebox{0.97}{$\omega_{{{{{\rm{BD}}}}}}$}}

\begin{document}

\title{
Post-Newtonian Feasibility  of  the Closed String Massless Sector 
}

\author{Kang-Sin Choi}
\email{kangsin@ewha.ac.kr}
\affiliation{Scranton Honors Program, Ewha Womans University,  Seodaemun-gu, Seoul 03760, Korea\\
Institute of Mathematical Sciences, Ewha Womans University,  Seodaemun-gu, Seoul 03760, Korea
}

\author{Jeong-Hyuck Park}
\email{park@sogang.ac.kr}
\affiliation{Department of Physics, Sogang University, 35 Baekbeom-ro, Mapo-gu, Seoul 04107,  Korea}

\begin{abstract}
\noindent  
We perform  post Newtonian  analysis of Double Field Theory as a test of string theory in gravitational sector against observations.    We identify the Eddington--Robertson--Schiff parameters  $\beta_{{{\rm{PPN}}}}$,   $\gamma_{{{\rm{PPN}}}}$  with the charges of  electric $H$-flux  and  dilaton respectively, and further  relate   them  to  stress-energy tensor. We show ${\beta_{{{\rm{PPN}}}}=1}$ from weak energy condition and argue that the observation of   $\gamma_{{{\rm{PPN}}}}\simeq 1$  signifies   the ultrarelativistic equation of state  in baryons, or the suppression of gluon condensate.
\end{abstract}

                             
\maketitle

\textit{Introduction.}---
What is the gravitational theory that string theory predicts? The conventional  answer has been  General Relativity (GR) on account of  a metric $g_{\mu\nu}$  appearing  in   closed string quantisation. However,   the metric is  only one segment of the closed string massless sector that   includes two additional fields,  a two-form potential $B_{\mu\nu}$ and a scalar dilaton $\phi$.  Given the fact that the $\ODD$ symmetry of T-duality~\cite{Buscher:1987sk,Buscher:1987qj,Giveon:1988tt,Duff:1989tf,Tseytlin:1990nb,
Tseytlin:1990va} transforms the trio to one another, it  may be  not unreasonable  to regard  the whole sector as gravitational.  This idea has come true in recent years. The $\ODD$ manifest formulation of supergravity~\cite{Siegel:1993xq,Siegel:1993th,Hull:2009mi,Hull:2009zb,
Hohm:2010jy,Hohm:2010pp}, dubbed  Double Field Theory (DFT),   has matured into \textit{stringy gravity}  which is  based on    a  beyond-Riemannian geometry~\cite{Jeon:2010rw,Jeon:2011cn,Jeon:2011vx,Park:2013mpa}  and has its own version of `Einstein  field equations' carrying $\ODD$  indices~\cite{Angus:2018mep},   
\be
G_{AB}=T_{AB}\,.
\label{EDFE0}
\ee
As off-shell and on-shell  conserved $\ODD$ tensors,  the LHS and RHS  of the equality  represent  stringy curvature and  matter respectively. Parametrising the fundamental variables of DFT  in terms of the trio $\{g,B,\phi\}$ (\textit{c.f.~}\cite{Lee:2013hma,Ko:2015rha,Morand:2017fnv,Berman:2019izh,Blair:2019qwi,Cho:2019ofr,Park:2020ixf,Gallegos:2020egk} for non-Riemannian alternatives),  the above single formula  is decomposed into
\be
\ba{rll}
R_{\mu\nu}+2\trd_{\mu}(\partial_{\nu}\phi)-\quarter H_{\mu\rho\sigma}H_{\nu}{}^{\rho\sigma}
&\!=&\! K_{(\mu\nu)}\,,\\
\half e^{2\phi}\trd^{\rho}\!\left(e^{-2\phi}H_{\rho\mu\nu}\right)&\!=&\! K_{[\mu\nu]}\,,\\
R+4\Box\phi-4\partial_{\mu}\phi\partial^{\mu}\phi-\textstyle{\frac{1}{12}}H_{\lambda\mu\nu}H^{\lambda\mu\nu}&\!=&\!\To\,.
\ea
\label{EDFE}
\ee
While  $K_{(\mu\nu)}, K_{[\mu\nu]}, \To$ are \textit{a priori} the  components of the DFT  stress-energy tensor $T_{AB}$ in (\ref{EDFE0}), each line of (\ref{EDFE}) may be identified---at least on-shell---as   the equation of motion of  $g_{\mu\nu}, B_{\mu\nu}$ and a DFT-dilaton $d$  absorbing $\sqrt{-g}$ as $d\equiv\phi-\half\ln\sqrt{-g}$. The $\ODD$  symmetry principle, once  taken as the working hypothesis,  fixes their   coupling to matters completely. For example,  when  coupled to  the Standard Model~\cite{Choi:2015bga},    spin-half fermions (diffeomorphically  half-unit-weighted) respond   to the   $H$-flux~\cite{Jeon:2011vx} and gauge bosons do so to the dilaton~\cite{Jeon:2011kp}:
\be
\!\int\!\!\rmd^{4}x~\bar{\psi}\big[i
\gamma^{\mu}{(D_{\mu}}{+\textstyle{\frac{1}{24}}}H_{\mu\nu\rho}\gamma^{\nu\rho})-m\big]\psi-\quarter e^{-2d}\Tr{(F_{\mu\nu}}{F^{\mu\nu})}\,,
\label{QCD}
\ee
which gives~\cite{Jeon:2011vx,Angus:2018mep}  $\To=\quarter \Tr(F_{\mu\nu}F^{\mu\nu})$ and
\be
\!K_{\mu\nu\!}=_{\!}\textstyle{\frac{-ie^{2d}}{4}}\bar{\psi}\big(_{\!}\gamma_{\mu}\vec{D}_{\nu}
{- {\cev{D}}_{\nu}}\gamma_{\mu} {+\quarter }H_{\nu\rho\sigma}\gamma_{\mu}{}^{\rho\sigma}\big)_{\!}\psi{+\half} \Tr(F_{\mu\rho}F_{\nu}{}^{\rho}).
\label{Kmn}
\ee
Having stated the above, to the best of our knowledge, the gravitational coupling in quantum field theories has never been   tested experimentally. All  the observations are based on the geodesic motion (and deviation)  of    point particles including  photon. The $\ODD$ symmetry  then  enforces  the  particles to couple   to  $g_{\mu\nu}$ only  in the usual manner, ${-m_{\ast}}\scalebox{1.2}{$\int$}\!\rmd\tau\,\sqrt{-g_{\mu\nu}\dot{y}^{\mu}\dot{y}^{\nu}}$,  which gives  $\To=0$ and
\be
K^{\mu\nu}(x)=
\half m_{\ast}{\dis{\int}}\!\rmd\tau~\dot{y}^{\mu}\dot{y}^{\nu\,}e^{2d}\delta^{(4)\!}\left(x-y(\tau)\right)\,.
\label{particleK}
\ee
That is to say~\cite{Ko:2016dxa},   the strong equivalence principle is saved in the string frame. While the sources can be stringy, the probes are still  point-like up to now.

Yet, given the augmented stress-energy tensor, the gravitational physics of  DFT  should be  richer than that of GR.   To develop some physical intuition on the governing equations~(\ref{EDFE}), we may  take the following three steps.  \textit{i)} The second formula dictates how  $K_{[\mu\nu]}$ shapes  the $H$-flux.  \textit{ii)} Subtracting the third from  the trace of the first, one solves for the dilaton:
\be
\Box\big(e^{-2\phi}\big)=\left(K_{\mu}{}^{\mu}-\To+\textstyle{\frac{1}{6}}{H_{\lambda\mu\nu}H^{\lambda\mu\nu}}\right)e^{-2\phi}\,.
\label{Boxphi}
\ee  
\textit{iii)} The  metric is lastly determined  by the first formula.

It is the purpose of the present Letter to address the feasibility of the closed string massless sector, or ${D=4}$ DFT as \textit{stringy gravity}  alternative to GR,  within the  parametrised post Newtonian (PPN) formalism~\cite{Will:1972zz,Damour:1992we,Will:2014kxa}.   Keeping  
astrophysical  systems in mind rather than the entire Universe, we shall neglect any large scale contribution to $\To$ such as a  (non-critical string) cosmological constant~\cite{Taylor:1988nw,Damour:1992kf,Damour:1993id,Damour:2002mi}. We  postulate the spacetime to be asymptotically flat and static. The  Eddington--Robertson--Schiff parameters $\betappn,\gammappn$ are then defined  in an isotropic coordinate system,
\be
\ba{lll}
\rd s^{2}&=&-\left(1-\frac{2M\GN}{\rr}+\frac{2\betappn (M\GN)^2}{\rr^{2}}+\cdots\right)\!\rd t^{2}\\
{}&{}&+\left(1+\frac{2\gammappn M\GN}{\rr}+\cdots\right)\rd x^{i}\rd x^{j}\delta_{ij}\,,
\ea
\label{PPN}
\ee
where $\rr=\sqrt{x^{i} x^{j}\delta_{ij}}$ is the isotropic radius and $\GN$ is  the Newtonian constant of gravitation (setting ``$\alpha_{{\rm{PPN}}}\equiv1$").

The current  stringent observational bounds for the solar gravity  are  $\gammappn-1=(2.1\pm 2.3)\times 10^{-5}$ (Shapiro time-delay by Cassini spacecraft)~\cite{Will:2014kxa,Bertotti:2003rm} and adopting this value $\betappn-1=(-4.1\pm 7.8)\times 10^{-5}$  or $(0.4\pm 2.4)\times 10^{-4}$ (perihelion shifts of Mercury or Mars)~\cite{Will:2014kxa,Mars}. Other observations include $4\betappn-\gammappn-3=(4.44\pm 4.5)\times 10^{-4}$ for the Earth gravity~\cite{Williams:2004qba} and $\gammappn=0.98\pm 0.07$ on galactic size scales~\cite{Bolton:2006yz}. These  all agree  with  the GR  prediction $\betappn=\gammappn=1$ and rule out various alternative  gravities~\cite{Wei-Tou:1972zhn,Weinberg:1972kfs,Misner:1973prb}.  For example, a class of scalar-tensor theories, notably  Brans--Dicke,  gives
\be
\scalebox{1.1}{$
\gammappn=\frac{1+\oBD}{2+\oBD}\,,$}
\label{gBD}
\ee 
and thus the  observations  compel the coupling constant $\oBD$  unnaturally  huge.   Truncating $K_{[\mu\nu]}$,  $\To$, and the $H $-flux consistently,   the leftover  components of (\ref{EDFE}) coincide with  the  Brans--Dicke   field equations for $\oBD=-1$.  This has been often  regarded  as a ``smoking gun'' to necessarily  eliminate  the string dilaton $\phi$. Below,  analysing  (\ref{EDFE})   with care, 
we show  that  spherically symmetric static  objects in stringy gravity (henceforth collectively ``stars" though applicable to the Earth)  have ${\betappn=1}$ and make  $\gammappn$  depend on     equation-of-state parameters. As a byproduct,   we shall  point out that the well-known relation~(\ref{gBD}) is an artifact of  restricting  to `pressureless' stars. It  can  generalise  to 
 ${\gammappn=3p/\rho}$ for   ${\oBD=-1}$, and thus becomes phenomenologically viable with ultrarelativistic pressure.\vspace{5pt}

\textit{Post Newtonian  Double Field Theory.}---Following the three steps \textit{i), ii), iii)} above, it is possible to solve the linearised  version of (\ref{EDFE}),  using  the Green function method with retarded time $t^{\prime}= t-|{\bf{x}-\bf{x^{\prime}}|}$~\cite{Cho:2019npq}, 
\be
\ba{rll}
H_{\lambda\mu\nu}(x)\!\!&\simeq&\!-\frac{1}{2\pi}{\dis{\int}}\!\rd^{3} x^{\prime}~\frac{\,3\partial^{\prime}_{[\lambda}K_{\mu\nu]}(x^{\prime})\,}{\left|\bf{x}-\bf{x^{\prime}}\right|}\,,\\
\phi(x)\!\!&\simeq&\!\frac{1}{8\pi}{\dis{\int}}\!\rd^{3} x^{\prime}~\frac{\left[ K_{\mu}{}^{\mu}-\To+{\frac{1}{6}}H_{\lambda\mu\nu}H^{\lambda\mu\nu}\right](x^{\prime})}{\left|\bf{x}-\bf{x^{\prime}}\right|}\,,\\
g_{\mu\nu}(x)\!\!&\simeq&\!\eta_{\mu\nu}+\frac{1}{2\pi}{\dis{\int}}\!\rd^{3} x^{\prime}~\frac{\left[ K_{(\mu\nu)}
{+{\frac{1}{4}}}{H_{\mu\rho\sigma}H_{\nu}{}^{\rho\sigma}}\right](x^{\prime})}{\left|\bf{x}-\bf{x^{\prime}}\right|}\,.
\ea
\label{linearSOL}
\ee
Mapping this to the isotropic coordinate system, we can read off $M\GN$ and $\gammappn$, but not $\betappn$ which would require higher order analysis. Instead, we turn to a known ${D=4}$ three-parameter exact solution to (\ref{EDFE})~\cite{Burgess:1994kq}. It has  vanishing sources (${T_{AB}=0}$) and  can be identified as the most  general,  asymptotically flat, static, and spherically symmetric  \textit{vacuum} geometry (${G_{AB}=0}$)~\cite{Ko:2016dxa}.  For our purposes,  we perform a radial coordinate transformation, $r_{\there}=\rr+\frac{a^{2}+b^{2}}{16\rr}+\frac{\alpha_{\there}-\beta_{\there}}{2}$  from \cite{Ko:2016dxa},  and  put  the geometry decisively   in the  isotropic (as well as often spherical)  coordinate system,
\be
\ba{rll}
e^{2\phi}&\!=&\!\gamma_{+}\!\left(\frac{4\rr-\sqrt{a^{2}+b^{2}}}{4\rr+\sqrt{a^{2}+b^{2}}}\right)^{\frac{2b}{\sqrt{a^{2}+b^{2}}}}+\gamma_{-}\!\left(\frac{4\rr+\sqrt{a^{2}+b^{2}}}{4\rr-\sqrt{a^{2}+b^{2}}}\right)^{\frac{2b}{\sqrt{a^{2}+b^{2}}}}\,,\\
H_{(3)}&\!=&\! h\sin\vartheta\,\rd t\wedge\rd\vartheta\wedge\rd\varphi=h\,\rd t\wedge\left(\dis{\frac{\epsilon_{ijk}x^{i\,}{\rd x^{j}\wedge\rd x^{k}}}{2\rr^3}}\right)\,,\\
\rd s^{2}&\!=&\! g_{tt}(\rr)\,\rd t^{2}+g_{\rr\rr}(\rr)\left(\rd\rr^{2}+\rr^{2}\rd\Omega^{2}\right)\,,
\ea
\label{vacuum}
\ee
where   $\rd\Omega^{2}=\rd\vartheta^{2}+\sin^{2}\vartheta\rd\varphi^{2}$,  and 
\[
\ba{l}
\,g_{tt}(\rr)=-e^{2\phi(r)}\left(\frac{4\rr-\sqrt{a^{2}+b^{2}}}{4\rr+\sqrt{a^{2}+b^{2}}}\right)^{\frac{2a}{\sqrt{a^{2}+b^{2}}}}\,,\\
g_{\rr\rr}(\rr)=e^{2\phi(r)}\left(\frac{4\rr+\sqrt{a^{2}+b^{2}}}{4\rr-\sqrt{a^{2}+b^{2}}}\right)^{\frac{2a}{\sqrt{a^{2}+b^{2}}}}
\left(1-\frac{a^{2}+b^{2}}{16\rr^{2}}\right)^{\!2}\,.
\ea
\]
The  three ``free" parameters are $\{a,b,h\,|\,b^{2}{\geq h^{2}}\}$,   which  set   $\gamma_{\pm}=\frac{1}{2}\big(1\pm\sqrt{1-h^2/b^2}\big)$. 

The geometry~(\ref{vacuum})  can be   identified as the outer geometry of  a  (stringy) compact star.  Crucially,   from the full Einstein  equations~(\ref{EDFE}),  it becomes   possible to   ascribe the  three parameters to the  star's stress-energy tensor or $\{K_{t}{}^{t},K_{r}{}^{r},{K_{\vartheta}{}^{\vartheta}=K_{\varphi}{}^{\varphi}},K_{tr},K_{[\vartheta\varphi]}\}$~\cite{Angus:2018mep}, such as
\be
a=\!\frac{1}{4\pi}\dis{\int}\!\rd^{3}x~e^{-2d}\left( K_{i}{}^{i}-K_{t}{}^{t}-\To +\textstyle{\frac{1}{6}}H_{ijk}H^{ijk}\right)\,.
\label{a} 
\ee
Inside the star, while  the  electric $H$-flux is persistently of the  rigid  form~(\ref{vacuum}),   the magnetic  flux can be  nontrivial,
\be
H^{\rr\vartheta\varphi}=-2 e^{2d}\dis{\int_{\rr}^{\rstar}}\rd\rr^{\prime}~e^{-2d}K^{[\vartheta\varphi]}\,,
\label{magneticH}
\ee
where the  upper limit of the integral  $\rstar$ denotes the star's finite radius.  
As implied by the first  of (\ref{EDFE}) for the isotropic metric, the electric and the magnetic fluxes should meet
\be
H_{t\vartheta\varphi}H_{\rr}{}^{\vartheta\varphi}=-2 K_{(t\rr)}
\label{emH}\,,
\ee
while the second of (\ref{EDFE}) gives ${K_{[tr]}=0}$.

Now,  from the $1/r$-expansion of the  outer geometry~(\ref{vacuum}), 
we  are able to identify   the Newtonian mass~\cite{Ko:2016dxa}, 
\be
M\GN=\half\big(a+b\sqrt{1-h^{2}/b^{2}}\big)\,,
\label{NM}
\ee
and    the  two post Newtonian parameters, 
\be
\ba{cc}
\textstyle{(\betappn{-1})(M\GN)^{2}=\!\frac{~h^{2}}{4}}\,,&\,
\textstyle{(\gammappn{-1})M\GN=-b\sqrt{1{-\frac{h^{2}}{b^{2}}}}}\,.
\ea
\label{bg}
\ee
Since   the inverse relations hold,
\[
\ba{c}
h=\pm 2\sqrt{\betappn{-1}}\,M\GN\,,\qquad a=(\gammappn+1)M\GN\,,\\ b=(1-\gammappn)\sqrt{1+4\frac{\betappn{-1}}{(\gammappn{-1})^{2}}}\,M\GN\,,
\ea
\label{abh}
\]
the outer  geometry~(\ref{vacuum}) is also   completely   determinable      by  the triple of $\{M\GN,\betappn,\gammappn\}$, such as
\be
\ba{rll}
{\phi}&\simeq& 
\frac{(\gammappn{-1})M\GN}{2\rr}+\frac{(\betappn{-1})(M\GN)^{2}}{\rr^{2}}\,,\\
H_{(3)}&=&\pm\sqrt{\betappn{-1}}\,M\GN\,\rd t\wedge\left(\dis{\frac{\epsilon_{ijk}x^{i\,}{\rd x^{j}\wedge\rd x^{k}}}{\rr^3}}\right)\,,\\
g_{\rr\rr}&\simeq&1+\frac{2\gammappn M\GN}{\rr}+\frac{\left(6\betappn+7\gammappn^{2}-7\right)(M\GN)^{2}}{\rr^{2}}\,.
\ea
\label{expansion}
\ee
The  expansion of $\phi$  suggests that    $(\gammappn{-1})M\GN$ is the \textit{dilaton charge}. Indeed, multiplying $\sqrt{-g}$ to (\ref{Boxphi}) and using $\sqrt{-g}\Box(e^{-2\phi})=\partial_{\mu}(\sqrt{-g}\partial^{\mu}e^{-2\phi})$, we can  compute  the dilaton charge  from the stress-energy tensor of the star,
\be
\ba{rll}
\!(\gammappn{-1})M\GN&\!\!=&\!\!\frac{1}{4\pi}{\dis{\oint_{\infty}}}\rd S_{i\,}\partial^{i} e^{-2\phi}\\
{}&\!\!=&\!\!\frac{1}{4\pi}\!\dis{\int}\!\rd^{3}x\,e^{-2d}\big(
K_{\mu}{}^{\mu}{-\To}{+\textstyle{\frac{1}{6}}}H_{\lambda\mu\nu}H^{\lambda\mu\nu}\big),
\ea
\label{gammaphi}
\ee
where the $H$-flux has been  fixed by  (\ref{magneticH}) and  (\ref{emH}). 
Similarly, from (\ref{expansion})  we identify $\sqrt{\betappn{-1}}M\GN$  as the  \textit{electric  $H$-flux charge} which is through    (\ref{magneticH}) and  (\ref{emH}) also related to the stress-energy tensor,
\be
\scalebox{0.98}{$
\sqrt{\betappn{-1}}M\GN\!=\left|{\dis{\oint_{\infty}}}\frac{\rd S_{i}}{16\pi}\epsilon^{ijk}H_{tjk}\right|\!
=\!\dis{\left|\frac{K_{(t\rr)}g^{\rr\rr}(e^{-2d\!}{/_{\!}\sin}\vartheta)}{2\int_{\rr}^{\rstar}\!\rd\rr^{\prime}(e^{-2d}K^{[\vartheta\varphi]})}\right|.}$}
\label{betaH}
\ee
For consistency,  the   fractional  quantity at the end of (\ref{betaH}) must be  independent of $r$, while  the $\vartheta$-dependency is  trivially  canceled since  $K_{[\vartheta\varphi]}\propto \sin\vartheta$, $K^{[\vartheta\varphi]}\propto1/\sin\vartheta$,  and
\be
e^{-2d}= e^{-2\phi(r)}\sqrt{-g_{tt}(r)g^{3}_{\rr\rr}(r)}\,r^{2}\sin\vartheta\,.
\label{measure}
\ee
Lastly,    from (\ref{a}), (\ref{bg}), (\ref{gammaphi}),  the Newtonian mass~(\ref{NM})    has its own integral expression~\cite{Angus:2018mep},
\be
M=\frac{1}{4\pi\GN}\dis{\int}\!\rd^{3}x~e^{-2d}\left(
-K_{t}{}^{t}-\textstyle{\frac{1}{2}}H_{t\mu\nu}H^{t\mu\nu}\right)\,.
\label{MASS}
\ee
Substituting   the expression of  $-K_{t}{}^{t}>0$ for particles~(\ref{particleK})  into (\ref{MASS}), we    identify   $m_{\ast}$   in terms of the  rest mass, 
\be
m_{\ast}
=8\pi\GN m\,,
\label{mastm}
\ee 
	and further confirm that, like in GR, the Newtonian mass $M$   is given by the sum of the  energy $m\dot{t}=m/\sqrt{1-v^{2}}$ rather than the rest mass. However, in contrast to GR,  it is  \textit{a priori} $-4K_{t}{}^{t}$  rather than the  conventional energy,  or $\propto -2K_{t}{}^{t}+\To$, that enters the mass formula.  {This difference is significant  for gauge bosons, in particular gluons:  as seen from (\ref{Kmn})} only the electric field  {$F_{ti}$}   contributes    to the mass, but the magnetic field  {$F_{ij}$}   does not.

Since ${e^{-2d}}$ is the rightful integral measure in DFT, all the   integral expressions above  are `proper', \textit{i.e.~}invariant under static (radial)  diffeomorphisms, which is rather notoriously  not the case  for the Schwarzschild mass in GR.  \newpage

\textit{${\betappn=1}$ from a weak energy condition.}---The exact expressions of  the Newtonian mass~(\ref{MASS})  and  the magnetic $H$-flux~(\ref{magneticH})  are  in good agreement   with the  linearised general solutions~(\ref{linearSOL}). The former~(\ref{MASS}) is  obvious and the latter~(\ref{magneticH})  is  due to  a shell theorem.  However, the electric $H$-flux given through (\ref{bg}) and  (\ref{betaH}) appears highly non-perturbative. In fact, for the linearised expression~(\ref{linearSOL}),  when $K_{[\mu\nu]}$ therein  is static and spherical,   we should have  $\partial_{t}K_{[\mu\nu]}=0$ and  $K_{[t i]}= x_{i} f(r)$  for some radial function $f(r)$. Consequently $\partial_{[t}K_{ij]}$ vanishes and   the linear version of the    electric $H$-flux must be trivial.  We resolve this  discrepancy   by arguing that, the electric $H$-flux in the exact solution should be trivial  too. With (\ref{measure}), its contribution to the Newtonian mass~(\ref{MASS}) is  an $1/r^2$  integral,  $\scalebox{1.1}{${\int}_{0}^{\infty}$}\rmd r\,e^{-2\phi}{h^{2}}/(\GN\sqrt{-g_{tt}g_{rr\,}}r^{2})$, which---provided  the geometry at the centre of the  star  is  non-singular---would diverge   unless ${h=0}$.  Since the  mass~$M$ is finite and   a weak energy condition $-K_{t}{}^{t}>0$ should hold,  the electric $H$-flux must be  inevitably    trivial.   Besides, the last expression of   (\ref{betaH}) ought to be independent of $r$.  In its small $r$-limit,  from  (\ref{measure})   the numerator   vanishes. Hence we arrive at   the same conclusion~${h=0}$.  Consequently,  from (\ref{emH}) ${K_{(tr)}=0}$ and  from (\ref{bg})  ${\betappn=1}$.  
\vspace{5pt}

\textit{${\gammappn\simeq  1}\,$ from  subhadronic pressure.}---With the vanishing  electric $H$-flux ({$h=0$}),  the volume integrals of the total mass~(\ref{MASS}) and  $\gammappn$~(\ref{gammaphi})  are now all restricted to the star's   interior,
\be
\gammappn=1{\,-}\scalebox{0.95}{$\,\frac{\dis{\int_{\rm{star}}}\!\!\!\!\!\!\rd^{3}x~e^{-2d}\big(K_{\mu}{}^{\mu}-\To+\textstyle{\frac{1}{6}}H_{ijk}H^{ijk}\big)}{\dis{\int_{\rm{star}}}\!\!\!\!\!\!\rd^{3}x~e^{-2d}K_{t}{}^{t}}$}\,,
\label{gammappn}
\ee
where ${\textstyle{\frac{1}{6}}H_{ijk}H^{ijk}=H_{r\vartheta\varphi}H^{r\vartheta\varphi}}$ is set by $K^{[\vartheta\varphi]}$ from (\ref{magneticH}).   Clearly, $\gammappn$ depends on the stress-energy tensor for which we  introduce volume-averaged,  equation-of-state parameters (\textit{c.f.~}\cite{Angus:2019bqs}),
\be
\ba{ll}
\scalebox{0.95}{$\wK=\frac{\dis{\int_{\rm{star}}}\!\!\!\!\!\!\rd^{3}x~e^{-2d\,}\textstyle{\frac{1}{3}}K_{i}{}^{i}}{\dis{\int_{\rm{star}}}\!\!\!\!\!\!\rd^{3}x~e^{-2d\,}(-K_{t}{}^{t})}\,,$}\quad&~~\scalebox{0.95}{$
\wT=\frac{\dis{\int_{\rm{star}}}\!\!\!\!\!\!\rd^{3}x~e^{-2d\,}\To}{\dis{\int_{\rm{star}}}\!\!\!\!\!\!\rd^{3}x~e^{-2d\,}(-K_{t}{}^{t})}\,,$}
\ea
\label{ww}
\ee
and further for the magnetic  $H$-flux part, 
\be\scalebox{0.95}{$
\deltaH\!=\!\frac{16\pi\dis{\int_{0}^{\rstar}}\!\!\!\rmd r~r^{2}e^{2\phi\!}\sqrt{g_{rr}^{3}/(-g_{tt})}
\Big(\!\int_{\rr}^{\rstar}\!\rd\rr^{\prime}\,e^{-2d}K^{[\vartheta\varphi]}\Big)^{\!2}}{
\dis{\int_{\rm{star}}}\!\!\!\!\!\!\rd^{3}x~e^{-2d\,}(-K_{t}{}^{t})}\,.$}
\label{deltaH}
\ee
Note from the identification of the usual energy-momentum tensor   $\cT_{\mu\nu}=e^{-2\phi}\big(2K_{(\mu\nu)}- g_{\mu\nu}\To\big)$ \textit{c.f.~}(\ref{CONSERVATION}),  the conventional equation-of-state  parameter reads
\be
w=\frac{~~\cT_{i}{}^{i}}{-3\cT_{t}{}^{t}}=\frac{p}{\rho}=\frac{\,\wK-\frac{1}{2}\wT\,}{1+\frac{1}{2}\wT}\,.
\ee
The post Newtonian parameter~(\ref{gammappn})  amounts to
\be
\gammappn=3\wK-\wT+\deltaH\,.
\label{wK2}
\ee
For the ideal gas of    particles, we have  $\wT=\deltaH=0$  and hence simply $\gammappn=3w$.  In terms of   the averaged speed $v$,   this is  equivalent to   $\gammappn\simeq  v^{2}$ 
since from  (\ref{particleK}) we have  ${K_{\mu}{}^{\mu}}/{K_{t}{}^{t}}=-1/({g_{tt}\dot{t}^{2}})=-g^{tt}(1-v^{2})\simeq 1-v^{2}$. 

If a star  were composed of non-relativistic pressureless  dust, we would get     ${\gammappn=0}$,  which certainly   fails  to explain the  observations. This is actually the case with the usual analysis of  the Brans--Dicke theory having  ${\oBD=-1}$~(\ref{gBD}) and  non-relativistic sources.    More  realistic stars would exert  pressure along  the  radial direction to balance  the gravitational force.  In the Newtonian gravity, the pressure is  fixed to meet $\frac{\rmd p(r)}{\rmd r}=-M(r)\GN\rho/r^{2}$ and leads to     $w=\frac{\langle p\rangle}{\rho}=\frac{M\GN}{2\rstar}$ (volume-averaged). The Sun would have   then merely  $\gammappn=3w\sim 3\times 10^{-6}$ which is far less than unity.  In GR,  the outer Schwarzschild geometry of a star  depends on the total mass only, being  inert to  the pressure. In DFT, the outer geometry~(\ref{vacuum})  depends on the equation-of-state  parameters~(\ref{ww}), yet the  Einstein equations thereof~(\ref{EDFE0}) still do not impose any restriction on them, as  seen \textit{e.g.~}from (\ref{gammaphi}) and  (\ref{MASS}).   In   flat spacetime,    the conservation of the (ordinary) stress-energy tensor implies that the volume integral of  the pressure  vanishes, which is known as the  von Laue  condition~\cite{Laue:1911lrk}. To correctly include the stringy gravitational effect, we call    the on-shell conservation $\na_{A}T^{A}{}_{B}=0$ of (\ref{EDFE0})~\cite{Angus:2018mep}~(\textit{c.f.~}\cite{Blair:2015eba,Park:2015bza})   which becomes more   concretly   $\trd_{\mu}\big(e^{-2\phi}K^{[\mu\nu]}\big)=0$ and 
\be
\trd^{\rho}\big(e^{-2\phi}K_{(\rho\mu)}\big)+\half e^{-2\phi}\big( H_{\mu\rho\sigma}K^{[\rho\sigma]}-\partial_{\mu}\To\big)=0\,.
\label{CONSERVATION}
\ee    
For the spherically symmetric static case of our interest where   $K_{tr}=0=H_{t\vartheta\varphi}$ and (\ref{magneticH})  holds,    writing   the covariant derivative in   (\ref{CONSERVATION})  explicitly as
\[
\ba{l}
r\sqrt{-g}\trd^{\rho}\big(e^{-2\phi}K_{(\rho r)}\big)
= \frac{\rmd~}{\rmd r}\Big[re^{-2d}\big(K_{r}{}^{r}-\frac{1}{3}K_{i}{}^{i}\big)\Big]\\
\quad+\frac{1}{3}r^3\frac{\rmd~}{\rmd r}\big(r^{-2}e^{-2d}K_{i}{}^{i}\big)
-\half re^{-2d}\big(\frac{\dot{g}_{tt}}{g_{tt}}K_{t}{}^{t}+\frac{\dot{g}_{rr}}{g_{rr}}K_{i}{}^{i}
\big)\,,
\ea
\]
we  relax   the   von Laue  condition rather inevitably,
\be
\ba{l}
{\dis{\int}\rd^{3}x}~e^{-2d}\big(K_{i}{}^{i}-\frac{3}{2}\To+\frac{1}{8} H_{ijk}H^{ijk}\big)\\
={\dis{\int}\rd^{3}x}~e^{-2d}\Big[\,
\frac{\,r\dot{g}_{tt}\,}{2g_{tt}}\big({-K_{t}{}^{t}}+\half\To+\frac{1}{24}H_{ijk}H^{ijk}\big)\\
-\frac{\,r\dot{g}_{rr}\,}{2g_{rr}}\big(K_{i}{}^{i}
{-\frac{3}{2}\To}{+\frac{1}{8}H_{ijk}}H^{ijk}\big)
{-r}\dot{\phi}\big(\To{+\frac{1}{12}}H_{ijk}H^{ijk}\big)
\Big].
\ea
\label{Laue}
\ee
While $\frac{\dot{g}_{tt}}{2g_{tt}}({-K_{t}{}^{t}})$  corresponds to the  gravitational  force, the presence of the  self-reflective term of  $-\frac{r\dot{g}_{rr}}{2g_{rr}}\big(K_{i}{}^{i}-\frac{3}{2}\To+\frac{1}{8}H_{ijk}H^{ijk})$  invalidates  the aforementioned   Newtonian  underestimation of the total  pressure. In fact, (\ref{CONSERVATION}) and (\ref{Laue}) are identities for any matter and  geometric data $\{K_{\mu\nu},\To,g_{\mu\nu},B_{\mu\nu},\phi\}$ that satisfy  the Einstein equations~(\ref{EDFE}),  since $\na_{A}G^{A}{}_{B}=0$ is a `Bianchi' identity in DFT~\cite{Hohm:2011si,Park:2015bza}.

For $\gammappn\simeq  v^{2}$ to be close to unity within the stringent observational bound  of $2.1\times 10^{-5}$~\cite{Will:2014kxa,Bertotti:2003rm}, the constituting particles---if not strings---should be ultrarelativistic. In terms of  temperature  $3\kB T={m}\big(\frac{1}{\sqrt{1-v^2}}-1\big)$,   we get $
\gammappn\simeq 1-\big(\frac{m}{m+3\kB T}\big)^{2}$. 
Assuming   $T=1.57\times 10^{7}\,\rm{K}$  at the  center of the Sun~\cite{NASA},  from $\big(\frac{m}{m+3\kB T}\big)^{2}< 2.1\times 10^{-5}$, the particle mass $m$ should not exceed $20\,\rm{eV}$. Certainly no such a  light  atom exists. While so,  a recent experiment has revealed  a    pressure  inside protons as high as   $10^{35}\,\rm{Pa} \sim0.005\,\rm{GeV}^{4}$~\cite{Burkert:2018bqq}  comparable to their mass density $10^{18}\,\rm{kg/m^3}\sim0.004\,\rm{GeV}^{4}\sim(\LambdaQCD)^{4}$~\cite{Ji:1994av}.  Henceforth we look into the energy and pressure  inside baryons, or the  QCD matter of (\ref{QCD}). The deviation $\gammappn{-1}$~(\ref{gammappn}) is then given  by  the sum of $\deltaH$~(\ref{deltaH}) and an  integral,
\be
\textstyle{\frac{1}{4\pi M\GN}\!}\dis{\int_{\rm{star}}}\!\!\!\!\!\!\!\rd^{3}x~
\quarter e^{-2d}\Tr(F_{\mu\nu}F^{\mu\nu})-\half \bar{\psi}(m+i\textstyle{\frac{1}{12}} \gamma^{ijk}H_{ijk})\psi\,.
\label{QCDK}
\ee
In our normalisation~(\ref{EDFE}),  from (\ref{mastm}),  $K_{\mu\nu},\To$ are order of $8\pi\GN$. Hence  the magnetic $H$-flux~(\ref{magneticH}) gives  a subleading  contribution and may be ignored.  Then,  (\ref{QCDK}) boils down  {faithfully} to {the integral of} the gluon and quark condensates {(vacuum expectation value)}~\cite{Vainshtein:1975sv,Shifman:1975tn}. According to  \cite{Casher:1974xd,Maris:1997hd,Maris:1997tm,Brodsky:2008xm,Brodsky:2008xu,Brodsky:2009zd,Brodsky:2010xf,Chang:2011mu,Brodsky:2012ku}, the condensates  have  local  support   restricted to the hadrons'  interior. Put differently,   the energy,   pressure, and condensates  specified  by the components of $K_{\mu\nu},\To$  are all confined.  Although   the  condensates may have the natural order of magnitudes  $(\LambdaQCD)^{4}$ and $m\times(\LambdaQCD)^{3}$ respectively, 
while the latter can be chirally rotated and averaged away,   the empirical measurements of the former~\cite{ALEPH:1996oqp,Geshkenbein:2001mn,Ioffe:2002be,Davier:2007ym,Davier:2008sk} 
 show significant scatter and even differences in sign~\cite{Brodsky:2008xm}, allowing  zero  as $0.006\pm0.012\,\rm{GeV}^{4}$~\cite{ALEPH:1996oqp}.  Perhaps, at  (four-dimensional)  thermal equilibrium,   the equal number of electric and magnetic {gluon} field-strength-squared cancel each other to suppress  the gluon condensate normalized by the baryon mass density, \textit{c.f.~}\cite{Brodsky:2009zd}. This ratio essentially  estimates (\ref{QCDK}) and  the observation  $\gammappn\simeq 1$ is consistent  with  the suppression.\vspace{5pt}

\textit{Discussion.}---To conclude,   DFT sets $\betappn=1$  and lets $\gammappn$ depend on the equation-of-state parameters~(\ref{wK2}).  Rather than ruling out the theory, applied to baryons' interior  where the energy and pressure are both confined, the apparently universal  observations   $\gammappn\simeq 1$  including  the Sun and  the Earth~\cite{Will:2014kxa,Bertotti:2003rm,Williams:2004qba}    signify   the ultrarelativistic equation of state inside baryons, ${\wK\simeq\frac{1}{3}},{\wT\simeq0}$,  through the suppression  of the {(massless)}  gluon condensate~(\ref{QCDK}).  
We call for both theoretical and experimental verifications of this  rather drastic conclusion of ordinary matter being  ultrarelativistic at a subhadronic level.

Some comments are in order. In terms of  the adiabatic index $c_{P}/c_{V}$,  one gets $w=c_{P}/c_{V}-1$ which becomes $1/3$ for ultrarelativistic ideal fluid having  $c_{P}=4 k_{{\rm{B}}}$ and $c_{V}=3k_{{\rm{B}}}$,  \textit{e.g.~}\cite{Park:2010fg}.  

Any  two-derivative potential-free  effective scalar theory,  if  $\ODD$-symmetric, should be of the form,
\be
S_{{\rm{eff.}}}={-\dis{\int\!}}\rmd^{4}x~e^{-2d}g^{\mu\nu}\partial_{\mu}\Phi^{I}\partial_{\nu}\Phi^{J}\cG_{IJ}(\Phi)\,,
\ee
which gives  $K_{\mu}{}^{\mu}=g^{\mu\nu}\partial_{\mu}\Phi^{I}\partial_{\nu}\Phi^{J}\cG_{IJ}=\To$, $K^{[\vartheta\varphi]}=0$, and thus  rather precisely $\gammappn=1$.

The $B$-field  is dual to a scalar axion in four dimensions. Recently,   a  static K\"{a}hler axion  has been  suggested to  make $\gammappn$ close to unity   (for ${w=0}$ and ${\oBD=\frac{3}{2}}$)~\cite{Burgess:2021qti}. In our case, the axion dual to the magnetic $H$-flux would be   time, rather than radial, dependent.

Since $\deltaH\geq 0$, if any star  were ever found to feature $\gammappn>1$, it should be stringy. From the superextensive  multi-integrals appearing  in $\deltaH$~(\ref{deltaH}),   bigger stringy stars or galaxies  would have larger value of $\gammappn$.  

The equations of state ${\wK\simeq\frac{1}{3}}$, ${\wT\simeq0}$ for $\gammappn\simeq1$ prevents also $\phi$ from  the cosmological time evolution~\cite{Tseytlin:1991xk,Angus:2019bqs}.   {This  may satisfy an extremely tight bound on  the time variation of  the fine structure constant~\cite{Webb:1998cq}, since $e^{-2d}$ is an overall factor hence affecting the coupling constant  of  the  gauge boson Lagrangian~(\ref{QCD}).}

\hfill

{\textit{Acknowledgments.}}---We wish to   thank    Robert Brandenberger, Minkyoo Kim,  Bum-Hoon Lee, Chang-Hwan Lee,  Hocheol Lee, Taekoon Lee, Wonwoo Lee, Shinji Mukohyama, Sang-Jin Sin,  and Clifford Will for useful comments and correspondences.  This work is supported by Basic Science Research Program through the National Research Foundation of Korea (NRF)  Grants, NRF-2016R1D1A1B01015196,  NRF-2018R1A2B2007163, and  NRF-2020R1A6A1A03047877 (Center for Quantum Space Time). 
\hfill

\begin{thebibliography}{99}
\bibitem{Buscher:1987sk}
  T.~H.~Buscher,
  ``A Symmetry of the String Background Field Equations,''
  Phys.\ Lett.\ B {\bf 194} (1987) 59.



\bibitem{Buscher:1987qj}
  T.~H.~Buscher,
  ``Path Integral Derivation of Quantum Duality in Nonlinear Sigma Models,''
  Phys.\ Lett.\ B {\bf 201} (1988) 466.




\bibitem{Giveon:1988tt}
A.~Giveon, E.~Rabinovici and G.~Veneziano,
``Duality in String Background Space,''
Nucl. Phys. B \textbf{322} (1989), 167-184
doi:10.1016/0550-3213(89)90489-6



\bibitem{Duff:1989tf}
M.~Duff,
``Duality Rotations in String Theory,''
Nucl. Phys. B \textbf{335} (1990), 610



\bibitem{Tseytlin:1990nb}
  A.~A.~Tseytlin,
  ``Duality Symmetric Formulation Of String World Sheet Dynamics,''
  Phys.\ Lett.\  B {\bf 242}, 163 (1990).

\bibitem{Tseytlin:1990va}
  A.~A.~Tseytlin,
  ``Duality Symmetric Closed String Theory And Interacting Chiral Scalars,''
  Nucl.\ Phys.\  B {\bf 350}, 395 (1991).
  




\bibitem{Siegel:1993xq}
  W.~Siegel,
  ``Two vierbein formalism for string inspired axionic gravity,''
  Phys.\ Rev.\ D {\bf 47} (1993) 5453
  [hep-th/9302036].

\bibitem{Siegel:1993th}
  W.~Siegel,
  ``Superspace duality in low-energy superstrings,''
  Phys.\ Rev.\ D {\bf 48} (1993) 2826
  [hep-th/9305073].
  
  
\bibitem{Hull:2009mi}
  C.~Hull and B.~Zwiebach,
  ``Double Field Theory,''
  JHEP {\bf 0909} (2009) 099
  [arXiv:0904.4664 [hep-th]].


\bibitem{Hull:2009zb}
  C.~Hull and B.~Zwiebach,
  JHEP {\bf 0909} (2009) 090
  [arXiv:0908.1792 [hep-th]].
  
  
  
  
\bibitem{Hohm:2010jy}
  O.~Hohm, C.~Hull and B.~Zwiebach,
  ``Background independent action for double field theory,''
  JHEP {\bf 1007} (2010) 016
  [arXiv:1003.5027 [hep-th]].
  
  
  
  


\bibitem{Hohm:2010pp}
O.~Hohm, C.~Hull and B.~Zwiebach,
``Generalized metric formulation of double field theory,''
JHEP \textbf{08} (2010), 008
[arXiv:1006.4823 [hep-th]].


  




\bibitem{Jeon:2010rw}
  I.~Jeon, K.~Lee and J.~H.~Park,
  ``Differential geometry with a projection: Application to double field theory,''
  JHEP {\bf 1104} (2011) 014
  doi:10.1007/JHEP04(2011)014
  [arXiv:1011.1324 [hep-th]].

\bibitem{Jeon:2011cn}
  I.~Jeon, K.~Lee and J.~H.~Park,
  ``Stringy differential geometry, beyond Riemann,''
  Phys.\ Rev.\ D {\bf 84} (2011) 044022
  doi:10.1103/PhysRevD.84.044022
  [arXiv:1105.6294 [hep-th]].


\bibitem{Jeon:2011vx}
I.~Jeon, K.~Lee and J.~H.~Park,
``Incorporation of fermions into double field theory,''
JHEP \textbf{11} (2011), 025
doi:10.1007/JHEP11(2011)025
[arXiv:1109.2035 [hep-th]].

\bibitem{Park:2013mpa}
J.~H.~Park,
``Comments on double field theory and diffeomorphisms,''
JHEP \textbf{06} (2013), 098
doi:10.1007/JHEP06(2013)098
[arXiv:1304.5946 [hep-th]].








\bibitem{Angus:2018mep}
  S.~Angus, K.~Cho and J.~H.~Park,
  ``Einstein Double Field Equations,''
  Eur.\ Phys.\ J.\ C {\bf 78} (2018) no.6,  500
  doi:10.1140/epjc/s10052-018-5982-y
  [arXiv:1804.00964 [hep-th]].  \\
We have mapped $G_{AB}=8\pi G_{\there} T_{AB}^{\there}$ of \cite{Angus:2018mep} to (\ref{EDFE0}) by  putting $16\pi G_{\there}=1$ and $\half T^{\there}_{AB}=T_{AB}$. The physically observable   Newton constant  $\GN$ appears through  the post Newtonian parametrisation of the metric~(\ref{PPN}).


\bibitem{Lee:2013hma}
  K.~Lee and J.~H.~Park,
  ``Covariant action for a string in doubled yet gauged  spacetime,''
  Nucl.\ Phys.\ B {\bf 880} (2014) 134 
  [arXiv:1307.8377 [hep-th]].
  



\bibitem{Ko:2015rha}
S.~M.~Ko, C.~Melby-Thompson, R.~Meyer and J.~H.~Park,
``Dynamics of Perturbations in Double Field Theory \& Non-Relativistic String Theory,''
JHEP \textbf{12} (2015), 144
doi:10.1007/JHEP12(2015)144
[arXiv:1508.01121 [hep-th]].

\bibitem{Morand:2017fnv}
  K.~Morand and J.~H.~Park,
  ``Classification of non-Riemannian doubled-yet-gauged spacetime,''
  Eur.\ Phys.\ J.\ C {\bf 77} (2017) no.10,  685
  [arXiv:1707.03713 [hep-th]].






\bibitem{Berman:2019izh}
D.~S.~Berman, C.~D.~A.~Blair and R.~Otsuki,
``Non-Riemannian geometry of M-theory,''
JHEP \textbf{07} (2019), 175
doi:10.1007/JHEP07(2019)175
[arXiv:1902.01867 [hep-th]].


\bibitem{Blair:2019qwi}
C.~D.~A.~Blair,
``A worldsheet supersymmetric Newton-Cartan string,''
JHEP \textbf{10} (2019), 266
doi:10.1007/JHEP10(2019)266
[arXiv:1908.00074 [hep-th]].


\bibitem{Cho:2019ofr}
K.~Cho and J.~H.~Park,
``Remarks on the non-Riemannian sector in Double Field Theory,''
Eur. Phys. J. C \textbf{80} (2020) no.2, 101
doi:10.1140/epjc/s10052-020-7648-9
[arXiv:1909.10711 [hep-th]].


\bibitem{Park:2020ixf}
J.~H.~Park and S.~Sugimoto,
``String Theory and non-Riemannian Geometry,''
Phys. Rev. Lett. \textbf{125} (2020) no.21, 211601
doi:10.1103/PhysRevLett.125.211601
[arXiv:2008.03084 [hep-th]].



\bibitem{Gallegos:2020egk}
A.~D.~Gallegos, U.~G\"ursoy, S.~Verma and N.~Zinnato,
``Non-Riemannian gravity actions from double field theory,''
[arXiv:2012.07765 [hep-th]].



\bibitem{Choi:2015bga}
K.~S.~Choi and J.~H.~Park,
``Standard Model as a Double Field Theory,''
Phys. Rev. Lett. \textbf{115} (2015) no.17, 171603
doi:10.1103/PhysRevLett.115.171603
[arXiv:1506.05277 [hep-th]].



\bibitem{Jeon:2011kp}
I.~Jeon, K.~Lee and J.~H.~Park,
``Double field formulation of Yang-Mills theory,''
Phys. Lett. B \textbf{701} (2011), 260-264
doi:10.1016/j.physletb.2011.05.051
[arXiv:1102.0419 [hep-th]].





\bibitem{Ko:2016dxa}
S.~M.~Ko, J.~H.~Park and M.~Suh,
``The rotation curve of a point particle in stringy gravity,''
JCAP \textbf{06} (2017), 002
doi:10.1088/1475-7516/2017/06/002
[arXiv:1606.09307 [hep-th]].










\bibitem{Will:1972zz}
C.~M.~Will and K.~Nordtvedt, Jr.,
``Conservation Laws and Preferred Frames in Relativistic Gravity. I. Preferred-Frame Theories and an Extended PPN Formalism,''
Astrophys. J. \textbf{177} (1972), 757
doi:10.1086/151754

\bibitem{Damour:1992we}
T.~Damour and G.~Esposito-Farese,
``Tensor multiscalar theories of gravitation,''
Class. Quant. Grav. \textbf{9} (1992), 2093-2176
doi:10.1088/0264-9381/9/9/015






\bibitem{Will:2014kxa}
C.~M.~Will,
``The Confrontation between General Relativity and Experiment,''
Living Rev. Rel. \textbf{17} (2014), 4
doi:10.12942/lrr-2014-4
[arXiv:1403.7377 [gr-qc]].




\bibitem{Taylor:1988nw}
T.~R.~Taylor and G.~Veneziano,
``Dilaton Couplings at Large Distances,''
Phys. Lett. B \textbf{213} (1988), 450-454
doi:10.1016/0370-2693(88)91290-7


\bibitem{Damour:1992kf}
T.~Damour and K.~Nordtvedt,
``General relativity as a cosmological attractor of tensor scalar theories,''
Phys. Rev. Lett. \textbf{70} (1993), 2217-2219
doi:10.1103/PhysRevLett.70.2217

\bibitem{Damour:1993id}
T.~Damour and K.~Nordtvedt,
``Tensor - scalar cosmological models and their relaxation toward general relativity,''
Phys. Rev. D \textbf{48} (1993), 3436-3450
doi:10.1103/PhysRevD.48.3436


\bibitem{Damour:2002mi}
T.~Damour, F.~Piazza and G.~Veneziano,
``Runaway dilaton and equivalence principle violations,''
Phys. Rev. Lett. \textbf{89} (2002), 081601
doi:10.1103/PhysRevLett.89.081601
[arXiv:gr-qc/0204094 [gr-qc]].

\bibitem{Bertotti:2003rm}
B.~Bertotti, L.~Iess and P.~Tortora,
``A test of general relativity using radio links with the Cassini spacecraft,''
Nature \textbf{425} (2003), 374-376
doi:10.1038/nature01997.


\bibitem{Mars}
A.~Konopliv, S.~Asmar, W.~Folkner, O.~Karatekin, D.~Nunes, S.~Smrekar, C.~Yoder and  M.~Zuber, ``Mars high resolution gravity fields from MRO, Mars seasonal gravity, and other dynamical parameters,'' 
Icarus \textbf{211} issue 1 (2011),  401-428 
doi:10.1016/j.icarus.2010.10.004.




\bibitem{Williams:2004qba}
J.~G.~Williams, S.~G.~Turyshev and D.~H.~Boggs,
``Progress in lunar laser ranging tests of relativistic gravity,''
Phys. Rev. Lett. \textbf{93} (2004), 261101
doi:10.1103/PhysRevLett.93.261101
[arXiv:gr-qc/0411113 [gr-qc]].


\bibitem{Bolton:2006yz}
A.~S.~Bolton, S.~Rappaport and S.~Burles,
``Constraint on the Post-Newtonian Parameter gamma on Galactic Size Scales,''
Phys. Rev. D \textbf{74} (2006), 061501(R)
doi:10.1103/PhysRevD.74.061501
[arXiv:astro-ph/0607657 [astro-ph]].


\bibitem{Wei-Tou:1972zhn}
N.~Wei-Tou,
``Theoretical frameworks for testing relativistic gravity. iv. a compendium of metric theories of gravity and their post-newtonian limits,''
Astrophys. J. \textbf{176} (1972), 769-796
doi:10.1086/151677


\bibitem{Weinberg:1972kfs}
S.~Weinberg,
``Gravitation and Cosmology: Principles and Applications of the General Theory of Relativity,'' 
John Wiley \& Sons (1972), 
ISBN: 9780471925675, 9780471925675 (Print).


\bibitem{Misner:1973prb}
C.~W.~Misner, K.~S.~Thorne and J.~A.~Wheeler,
``Gravitation,''
 Princeton University Press (1973),
ISBN: 0716703440, 9780716703440.





\bibitem{Burgess:1994kq}
C.~P.~Burgess, R.~C.~Myers and F.~Quevedo,
``On spherically symmetric string solutions in four-dimensions,''
Nucl. Phys. B \textbf{442} (1995), 75-96
doi:10.1016/S0550-3213(95)00090-9
[arXiv:hep-th/9410142 [hep-th]].





\bibitem{Cho:2019npq}
K.~Cho, K.~Morand and J.~H.~Park,
``Stringy Newton Gravity with $H$-flux,''
Phys. Rev. D \textbf{101} (2020) no.6, 064020
doi:10.1103/PhysRevD.101.064020
[arXiv:1912.13220 [hep-th]].


  

\bibitem{Angus:2019bqs}
S.~Angus, K.~Cho, G.~Franzmann, S.~Mukohyama and J.~H.~Park,
``$\mathbf {O}(D,D)$ completion of the Friedmann equations,''
Eur. Phys. J. C \textbf{80} (2020) no.9, 830
doi:10.1140/epjc/s10052-020-8379-7
[arXiv:1905.03620 [hep-th]].



\bibitem{Laue:1911lrk}
M.~Laue,
``Zur Dynamik der Relativit\"{a}tstheorie,''
Annalen Phys. \textbf{340} (1911) no.8, 524-542
doi:10.1002/andp.19113400808



\bibitem{Blair:2015eba}
C.~D.~A.~Blair,
``Conserved Currents of Double Field Theory,''
JHEP \textbf{04} (2016), 180
doi:10.1007/JHEP04(2016)180
[arXiv:1507.07541 [hep-th]].



\bibitem{Park:2015bza}
  J.~H.~Park, S.~J.~Rey, W.~Rim and Y.~Sakatani,
  ``$\ODD$ covariant Noether currents and global charges in double field theory,''
  JHEP {\bf 1511} (2015) 131
  [arXiv:1507.07545 [hep-th]].


\bibitem{Hohm:2011si}
O.~Hohm and B.~Zwiebach,
``On the Riemann Tensor in Double Field Theory,''
JHEP \textbf{05} (2012), 126
doi:10.1007/JHEP05(2012)126
[arXiv:1112.5296 [hep-th]].

\bibitem{NASA}
``NASA/Marshall Solar Physics", Marshall Space Flight Center  (2007), 
\url{https://solarscience.msfc.nasa.gov/}.


\bibitem{Burkert:2018bqq}
V.~D.~Burkert, L.~Elouadrhiri and F.~X.~Girod,
``The pressure distribution inside the proton,''
Nature \textbf{557} (2018) no.7705, 396-399
doi:10.1038/s41586-018-0060-z



\bibitem{Ji:1994av}
X.~D.~Ji,
``A QCD analysis of the mass structure of the nucleon,''
Phys. Rev. Lett. \textbf{74} (1995), 1071-1074
doi:10.1103/PhysRevLett.74.1071
[arXiv:hep-ph/9410274 [hep-ph]].

\bibitem{Vainshtein:1975sv}
A.~I.~Vainshtein, V.~I.~Zakharov and M.~A.~Shifman,
``A Possible mechanism for the Delta T = 1/2 rule in nonleptonic decays of strange particles,''
JETP Lett. \textbf{22} (1975), 55-56

\bibitem{Shifman:1975tn}
M.~A.~Shifman, A.~I.~Vainshtein and V.~I.~Zakharov,
``Light Quarks and the Origin of the Delta I = 1/2 Rule in the Nonleptonic Decays of Strange Particles,''
Nucl. Phys. B \textbf{120} (1977), 316-324
doi:10.1016/0550-3213(77)90046-3

\bibitem{Casher:1974xd}
A.~Casher and L.~Susskind,
``Chiral magnetism (or magnetohadrochironics),''
Phys. Rev. D \textbf{9} (1974), 436-460
doi:10.1103/PhysRevD.9.436


\bibitem{Maris:1997hd}
P.~Maris, C.~D.~Roberts and P.~C.~Tandy,
``Pion mass and decay constant,''
Phys. Lett. B \textbf{420} (1998), 267-273
doi:10.1016/S0370-2693(97)01535-9
[arXiv:nucl-th/9707003 [nucl-th]].

\bibitem{Maris:1997tm}
P.~Maris and C.~D.~Roberts,
``Pi- and K meson Bethe-Salpeter amplitudes,''
Phys. Rev. C \textbf{56} (1997), 3369-3383
doi:10.1103/PhysRevC.56.3369
[arXiv:nucl-th/9708029 [nucl-th]].

\bibitem{Brodsky:2008xm}
S.~J.~Brodsky and R.~Shrock,
``On Condensates in Strongly Coupled Gauge Theories,''
[arXiv:0803.2541 [hep-th]].

\bibitem{Brodsky:2008xu}
S.~J.~Brodsky and R.~Shrock,
``Standard-Model Condensates and the Cosmological Constant,''
Science \textbf{108} (2011), 45-50
[arXiv:0803.2554 [hep-th]].



\bibitem{Brodsky:2009zd}
S.~J.~Brodsky and R.~Shrock,
``Condensates in Quantum Chromodynamics and the Cosmological Constant,''
Proc. Nat. Acad. Sci. \textbf{108} (2011), 45-50
doi:10.1073/pnas.1010113107
[arXiv:0905.1151 [hep-th]].

\bibitem{Brodsky:2010xf}
S.~J.~Brodsky, C.~D.~Roberts, R.~Shrock and P.~C.~Tandy,
``Essence of the vacuum quark condensate,''
Phys. Rev. C \textbf{82} (2010), 022201(R) 
doi:10.1103/PhysRevC.82.022201
[arXiv:1005.4610 [nucl-th]].


\bibitem{Chang:2011mu}
L.~Chang, C.~D.~Roberts and P.~C.~Tandy,
``Expanding the concept of in-hadron condensates,''
Phys. Rev. C \textbf{85} (2012), 012201(R) 
doi:10.1103/PhysRevC.85.012201
[arXiv:1109.2903 [nucl-th]].


\bibitem{Brodsky:2012ku}
S.~J.~Brodsky, C.~D.~Roberts, R.~Shrock and P.~C.~Tandy,
``Confinement contains condensates,''
Phys. Rev. C \textbf{85} (2012), 065202
doi:10.1103/PhysRevC.85.065202
[arXiv:1202.2376 [nucl-th]].






\bibitem{ALEPH:1996oqp}
R.~Barate \textit{et al.} [ALEPH],
``Studies of quantum chromodynamics with the ALEPH detector,''
Phys. Rept. \textbf{294} (1998), 1-165
doi:10.1016/S0370-1573(97)00045-8


\bibitem{Geshkenbein:2001mn}
B.~V.~Geshkenbein, B.~L.~Ioffe and K.~N.~Zybablyuk,
``The Check of QCD based on the tau - decay data analysis in the complex q**2 - plane,''
Phys. Rev. D \textbf{64} (2001), 093009
doi:10.1103/PhysRevD.64.093009
[arXiv:hep-ph/0104048 [hep-ph]].


\bibitem{Ioffe:2002be}
B.~L.~Ioffe and K.~N.~Zyablyuk,
``Gluon condensate in charmonium sum rules with three loop corrections,''
Eur. Phys. J. C \textbf{27} (2003), 229-241
doi:10.1140/epjc/s2002-01099-8
[arXiv:hep-ph/0207183 [hep-ph]].


\bibitem{Davier:2007ym}
M.~Davier, A.~Hocker and Z.~Zhang,
``ALEPH Tau Spectral Functions and QCD,''
Nucl. Phys. B Proc. Suppl. \textbf{169} (2007), 22-35
doi:10.1016/j.nuclphysbps.2007.02.109
[arXiv:hep-ph/0701170 [hep-ph]].


\bibitem{Davier:2008sk}
M.~Davier, S.~Descotes-Genon, A.~Hocker, B.~Malaescu and Z.~Zhang,
``The Determination of alpha(s) from Tau Decays Revisited,''
Eur. Phys. J. C \textbf{56} (2008), 305-322
doi:10.1140/epjc/s10052-008-0666-7
[arXiv:0803.0979 [hep-ph]].







\bibitem{Park:2010fg}
J.~H.~Park and S.~W.~Kim,
``Existence of a critical point in the phase diagram of the ideal relativistic neutral Bose gas,''
New J. Phys. \textbf{13} (2011), 033003
doi:10.1088/1367-2630/13/3/033003
[arXiv:1001.1823 [cond-mat.quant-gas]].



\bibitem{Burgess:2021qti}
C.~P.~Burgess and F.~Quevedo,
``Axion Homeopathy: Screening Dilaton Interactions,''
[arXiv:2110.10352 [hep-th]].

\bibitem{Tseytlin:1991xk}
A.~A.~Tseytlin and C.~Vafa,
``Elements of string cosmology,''
Nucl. Phys. B \textbf{372} (1992), 443-466
doi:10.1016/0550-3213(92)90327-8
[arXiv:hep-th/9109048 [hep-th]].

\bibitem{Webb:1998cq}
J.~K.~Webb, V.~V.~Flambaum, C.~W.~Churchill, M.~J.~Drinkwater and J.~D.~Barrow,
``Evidence for time variation of the fine structure constant,''
Phys. Rev. Lett. \textbf{82} (1999), 884-887
doi:10.1103/PhysRevLett.82.884
[arXiv:astro-ph/9803165 [astro-ph]].
\end{thebibliography}
\end{document}